# Photon arrival time tagging with many channels, sub-nanosecond deadtime, very high throuphput, and fiber optic remote synchronization

Michael Wahl [a),1], Tino Röhlicke [a)], Sebastian Kulisch [a)], Sumeet Rohilla [a,b)], Benedikt Krämer [a)] and Andreas C. Hocke [b)]

a) PicoQuant GmbH, Rudower Chaussee 29, D-12489 Berlin, Germany
b) Charité – Universitätsmedizin Berlin, corporate member of Freie Universität Berlin, Humboldt-Universität zu Berlin, and Berlin Institute of Health, Department of Internal Medicine/Infectious Diseases and Respiratory Medicine, Charitéplatz 1, D-10117 Berlin, Germany

**Abstract**

Time-Correlated Single Photon Counting (TCSPC) and time tagging of individual photon detections are powerful tools in many quantum optical experiments and other areas of applied physics. Using TCSPC, e.g., for the purpose of fluorescence lifetime measurements, is often limited in speed due to dead-time losses and pile-up. We show that this limitation can be lifted by reducing the dead-time of the timing electronics to the absolute minimum imposed by the speed of the detector signals while maintaining high temporal resolution. A complementing approach to speedy data acquisition is parallelization by means of simultaneous readout of many detector channels. This puts high demands on the data throughput of the TCSPC system, especially in time tagging of individual photon arrivals. Here, we present a new design approach, supporting up to 16 input channels, an extremely short dead-time of 650 ps, very high time tagging throughput, and a timing resolution of 80 ps. In order to facilitate remote synchronization of multiple such instruments with highest precision, the new TCSPC electronics provide an interface for White Rabbit fiber optic networks. Beside fundamental research in the field of astronomy, such remote synchronization tasks arise routinely in quantum communication networks with node to node distances on the order of tens of kilometers. In addition to showing design features and benchmark results of new TCSPC electronics, we present application results from spectrally resolved and high-speed fluorescence lifetime imaging in medical research. We furthermore show how pulse-pile-up occurring in the detector signals at high photon flux can be corrected for and how this data acquisition scheme performs in terms of accuracy and efficiency.

[1] Author to whom correspondence should be addressed; electronic mail: wahl@picoquant.com

# I. INTRODUCTION

The method of Time-Correlated Single Photon Counting (TCSPC) is based on the precisely timed registration of single photons, e.g., of a fluorescence signal. Historically, the main application of TCSPC was the measurement of fluorescence lifetimes, which were usually probed via periodic excitation by a short flash or laser pulse. When registering an optical signal, a single-photon detector generates an electrical signal which is highly correlated with the photon arrival time at the detector. The reference for this photon timing signal is set by the corresponding excitation pulse often referred to as synchronization or sync signal. The TCSPC electronics must then record each photon timing signal with great precision relative to the corresponding synchronization event. A histogram of the resulting time differences reflects the fluorescence decay of the observed molecule or ensemble.[1]

The time difference measurement in TCSPC is obtained by means of fast electronics. High demands on time resolution exist in fluorescence lifetime measurements of organic dyes used for labeling biological macromolecules. Here, the required time resolution can be on the order of a few picoseconds. State-of-the-art time measurement circuits, so-called Time to Digital Converters (TDC), are capable of providing the required time resolution.[2]

For lifetime measurements, the histogramming of above-mentioned time differences is often implemented directly in the measurement electronics as a very resource-efficient data collection mode. In this TCSPC histogramming mode, the temporal ordering of events on timescales longer than the excitation period is discarded. However, keeping the temporal ordering of recorded detection-events gives access to a new class of measurements – temporal correlation measurements. Fluorescence Correlation Spectroscopy (FCS) is a prime example for such a method that exploits the fluctuations in photon arrival times over longer time scales, typically microseconds to milliseconds. From the fluorescence intensity fluctuations generated by molecules diffusing through a confocal volume, one can obtain information on, e.g., the translational diffusion constant and the number of molecules in the observed volume.[3] At the time of their inception, such measurements were typically implemented with hardware correlators. Today a TCSPC approach is often favored as the flexibility of time scales and the ability to evaluate the data with a variety of schemes provides tremendous benefits for the user.

Another correlation technique, the so-called second-order correlation or $g^{(2)}$ measurement, has become an essential method to gain knowledge about the photon statistics of the light field under observation. It is often implemented with a Hanbury-Brown and Twiss setup[4], which splits a light field via a 50/50 beam splitter and places two single-photon detectors at its output ports. If a sufficiently low number of emitters generate the light field, it reveals features on the picosecond to nanosecond time scale[5,6], namely a dip at zero delay between the detectors. This dip is interpreted as photon antibunching and can be used to determine the number of emitters in a molecular complex.[7-10] These techniques require a time resolution similar to classic TCSPC and cannot be performed with typical hardware correlators built for FCS measurements.

Over the last decades, quantum technology has matured and today a lot of effort is focused on putting it to use in real-world applications. In the case of optical quantum technologies, the generation of suitable states of light via single-photon sources is one of the most important engineering tasks at hand[11,12]. Here, the second order correlation measurement serves a new role – the depth of the dip for zero delay is a direct indicator for how well such a single photon source performs. For a single-photon state, $g^{(2)}(0)$ should be zero and recent solid-state sources come impressively close to this value[12,13]. On the other hand, these correlation measurements lie also at the heart of experiments in fundamental quantum mechanics,[14] e.g., the ability to extract even higher order correlations[15] from photon arrival times is routinely exploited.

In summary, the requirements of virtually all of these experimental techniques based on single photon arrival times are very similar and valuable information can be found on very different time scales. A first step towards unified instrumentation permitting all these experiments was a modification of classic TCSPC electronics. The start-stop timing circuitry was used as previously, providing the required picosecond time resolution for TCSPC. In order to maintain the information embedded in the temporal patterns of photon arrivals, the events were no longer stored as histograms, but as separate records. In addition, a coarser timing was performed on each photon event with respect to the start of the experiment.[16] This is referred to as time-tagged time-resolved (TTTR) data collection or more generically just "time tagging".

The TTTR concept avoids both redundancy in the data stream and loss of information. As a result, virtually all algorithms and methods for the analysis of photon dynamics can be implemented. For instance, intensity traces over time, as traditionally obtained from multi-channel scalers (MCS), are obtained from TTTR data by evaluating only the time tags of the photon records. This provides access to fluorescence bursts from freely diffusing single molecules or to blinking dynamics.[17,18] The time-tagged data format also allows for the combination of arrival timing with the detector channel information.[19] This combined information proves to be very powerful for the investigation of molecular dynamics, e.g., protein interaction.[20]

A great advantage of off-line analysis of time-tagged photon event data is that the type of analysis does not have to be defined at the time of measurement. The benefit, e.g., for FCS, is evident: Traditional hardware correlators perform an immediate (real-time) data reduction that does not conserve the original data, and thus excludes alternative types of data analysis. By having individual photon records available, one can perform the correlation using suitable software and select from a wide range of analysis methods without the need for additional measurements. On modern computers and using fast algorithms, it is also possible to perform the correlation in real-time.[21] By using the information of TTTR-type data, subpopulation-selective[22] or time-gated[23] intensity correlations can be calculated, and different molecular species in a mixture can be separated from a single FCS measurement. By weighting the photon events according to their TCSPC time in the correlation procedure, one can obtain the separate FCS curves for each species.[24] Another application of time-tagged TCSPC is fluorescence lifetime imaging (FLIM), for which the spatial origin of the photons must be recorded in addition to the TCSPC data. FLIM systems using on-board histogramming impose limits in the recordable image size. To avoid this, we previously extended the TTTR data stream concept to contain markers for synchronization information from a scanner[25], allowing the reconstruction of 2D or 3D images from the stream of TTTR records. The data is nearly free of redundancy and can therefore be transferred in real-time, even if the scan speed is as fast as in Laser Scanning Microscopes (LSM).[26]

In the original TTTR approach, the different time scales are processed and used independently. However, it is often of great interest to obtain high resolution timing on the overall scale, i.e. by combining coarse and fine timing into one global arrival time per event with picosecond resolution. It has been demonstrated that it is possible to combine coarse and fine timing in order to perform temporal analysis of single molecule fluorescence from the picosecond to second time scale.[27] In a more generalized approach, without assumptions on start and stop events, one collects precise time tags of all events of interest (excitation, emission or others). It is then possible to perform the desired analysis on the original event times, thereby covering almost all dynamic effects of the photophysics and other dynamics of fluorescent molecules.

The same high-resolution global arrival time tagging of photon detections is very valuable in quantum optical experiments. Ideally, correlation measurements are done on independent timing channels such that dead time effects can be eliminated by cross-correlating the detector signals.[28] In quantum communication,[29] the signal to noise ratio can be improved drastically by identifying background processes in the time tag analysis. The characterization of single photon sources,[30] which play a major role in quantum optics experiments, also benefits from independent, time tagged detection channels.[31]

## II. INSTRUMENT DESIGN

### A. New objectives

A limitation of virtually all existing high resolution TCSPC timing electronics is their dead time, i.e. the time needed to process a photon arrival before being ready for the next event. This causes losses of photon detections and distortion of the arrival time statistics. Although these limitations have been well known for a long time, until recently it has not been a top priority to reduce the dead time of the timing electronics. This was due to two reasons: first, many detectors also impose a considerable dead time in the presence of which a reduced dead time of the timing electronics would be in vain; second, the means of precise timing with an accuracy of a few picoseconds and dead times below some tens of nanoseconds were limited and would typically require a trade-off between the two figures of merit. However, recent detector developments, notably the emergence of Hybrid Photodetectors (HPD) and Superconducting Nanowire Single Photon Detectors (SNSPD) brought about interesting solutions with very good time resolution and short dead time.[32, 33] Simultaneously, modern high speed electronic components have been developed, making it possible to also significantly reduce the dead time of the timing electronics while maintaining a reasonably good resolution, matching that of the detectors.

Another frequent requirement regarding the timing electronics is a large number of independent input channels. In order to work with signals from multiple detectors (e.g., multiple wavelength channels) or from detector arrays[34-36], large numbers of inputs ranging from a few to several tens are of practical interest. Existing commercial solutions that multiplex several detector signals into one timing input rely on a low probability of receiving photons on more than one channel at the same time.[37] Any such simultaneous multi-hits must be discarded since they cannot be assigned unambiguously to one channel. The multiplexed channels are therefore not independent, which causes artifacts in correlation measurements. While the most prominent artifact is a gap in the correlation curves at lag times shorter than the dead time, more subtle and easily overlooked artifacts result from "blinding" other channels when one channel is subjected to high photon rates. For similar reasons, a multiplexing architecture forbids applications with high count rates, such as high-speed quantum communication, optical tomography or fast FLIM with multiple color channels.

Some existing solutions providing multiple truly independent channels are operating several TCSPC units in parallel, where each unit has one photon input channel and one synchronization input channel.[38] This requires fanning out the synchronization signal to each unit. However, when performing time tagging data collection, problems arise in assigning simultaneous events from two or more channels to the correct sync period due to the inevitable tolerances in sync timing. An ideal solution should provide many independent photon timing channels, all operating against one common synchronization input channel. Furthermore, parallel operation of complete TCSPC units, each with their own host computer interface, will result in multiple data streams arriving at the host computer. Relative timing between events across channels is then not immediately possible. This will cause complications in real-time analysis of time-tagged data where the order and temporal relation of events across all channels is critical. Ideally, the instrument should deliver a single data stream with all events from all channels in correct temporal order.

**B. Design features**

In order to meet the requirements outlined above, a new time tagging scheme has been designed with focus on shortest possible dead time. A very short dead time of only 650 ps has been achieved by using recent Field Programmable Gate Array (FPGA) technology and giving up some paradigms, notably that of highest possible time resolution. In comparison to previous instruments [28,39], this included first of all being satisfied with a time resolution of 80 ps, which is still reasonably matched to detectors of interest, notably those with a short dead time. A second change of paradigm was the omission of constant fraction discriminators (CFDs) at the inputs. Although these circuits are valuable in order to eliminate timing errors due to fluctuating detector pulse heights, they had to be given up because of them requiring a relatively long internal processing delay and the resulting dead time of at least some 10 ns. Indeed, this decision is not overly harmful for an instrument with moderate resolution since the improvement achievable with CFDs is typically quite small and only worthwhile for very high resolution instruments. In our particular case the detector of topmost interest is a HPD (PMA Hybrid Series, PicoQuant, Germany) with a transit time spread on the order of 100 ps, a transition time of 400 ps and a pulse height fluctuation of about 10 % (single photon response). The timing error due to pulse height fluctuation the CFD might correct for is therefore smaller than the transit time spread and the TDC resolution.

The chosen new system architecture is laid out to be scalable in terms of the number of input channels, while using one common synchronization channel. Furthermore, it is designed to generate a single time tagging data stream, delivering the time tag records for all events from all inputs in correct temporal order. Another key objective was achieving high throughput and uninterrupted real-time recording of this data stream at full resolution of the timing circuits, even at very high event rates.

The internal data formats and processing logic are laid out for scalability to a maximum of 64 channels. The first physical implementation of the housing is laid out to accommodate either 4 or 8 photon timing channels and a separate synchronization channel. The next larger housing is designed to accommodate 16 detector channels without any change in data formats or processing logic.

In addition to the regular timing inputs, the TCSPC electronics provide four additional inputs for TTL signals that are captured at lower resolution and get inserted in the data stream exactly like photon events. These low-resolution signals can serve as markers for different types of synchronization, e.g., representing spatial information of scanning devices.

A central crystal clock ensures that all timing inputs have a common time base. Optionally, the clock may also be fed in as an industry standard 10 MHz signal from an external source such as an atomic clock. The same type of 10 MHz standard clock signal is also available as an output so that multiple devices can be synchronized.

Another unique new feature relating to synchronization is the implementation of a White Rabbit interface. White Rabbit is a hardware and software protocol that enhances Gbit fiber-based Ethernet to carry accurate clock and time information across the network nodes over tens of kilometers.[40] For the WR-Protocol a precision of a few tens of picoseconds has been routinely achieved in numerous implementations over varying link-distances.[41] Such a precision is perfectly sufficient to match the new instrument's own timing resolution of 80 ps.

Given the very short dead time $T_d$ of only 650 ps, it is in principle possible that there can be photon detections at a rate of up to $1/T_d = 1.53$ GHz in each channel. Considering that the design is aimed at many parallel channels, it is clear that the chain of time tag data processing and the transfer over USB cannot sustain the full theoretical rate. However, this is not necessary. Since the photon arrivals are of statistical nature it is typically sufficient to be able to handle short bursts of photon arrivals while the average rate is significantly smaller. The input timing units therefore contain FiFo (First in First out) buffers for 2048 event records. These take care of data that must temporarily wait to be transferred to the main processing unit via high speed serial links. Using serial links here is an important prerequisite for the desired scalability of the system at reasonable cost. The main processing unit also contains FiFo buffers for the event records coming from the serial links. The FiFo buffers are important for the desired continuous and uninterrupted operation at high event rates. However, they introduce a temporal decoupling of the event records. Since the event rates on individual input channels can be extremely different (or even zero), the time-tagged event records delivered to the main processing unit may arrive unpredictably disordered. The actual arrival time of the records is no longer in accordance with the original event times. However, such an order is required in the data stream to the host computer because the latter is usually processing this data in real-time. The processing algorithms are typically very demanding so that ordered data is a strict requirement. The present design uses a previously introduced sorter scheme that solves this issue.[39]

The main processing unit also comprises a FiFo buffer for the sorted event records to be sent to the host computer interface. This ensures that sorting and other processing is decoupled from the temporal structure of the host transfer process. Furthermore, the main processing unit comprises an overall measurement and data flow controller that takes commands from the host computer and signals any conditions the host may need to be informed of. The processing performed in the timing units and in the main processing units is implemented in FPGA.

Figure 1 shows a block diagram of the new time tagging architecture. All timing inputs have a programmable trigger level and edge (rising/falling).

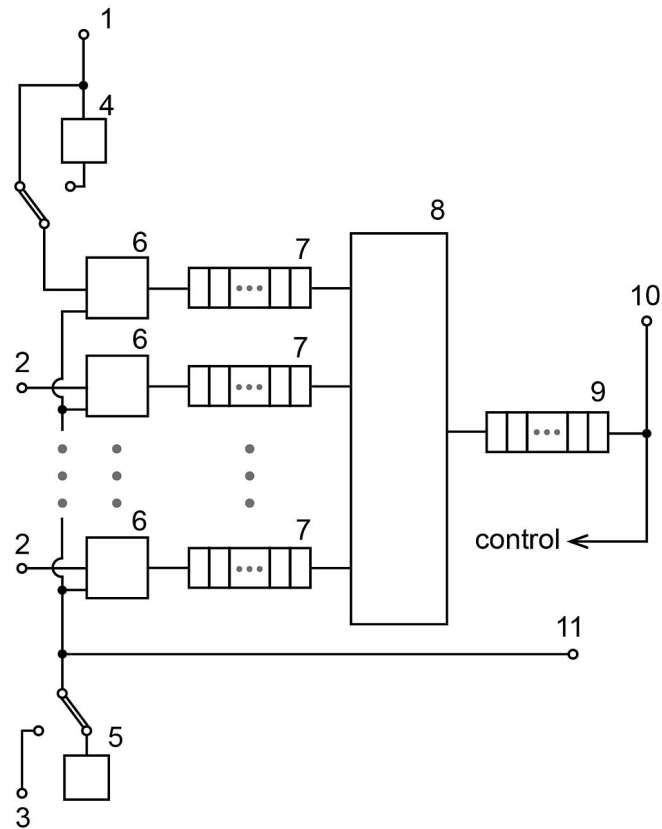

| | | |
|---|---|---|
| 1 | | Synchronization input |
| 2 | | Timing input |
| 3 | | External reference clock input |
| 4 | | Sync divider with bypass |
| 5 | | Internal crystal clock oscillator |
| 6 | | TDC |
| 7 | | Fifo buffer for 2048 events @ full rate |
| 8 | | Sorter or histogrammer @ 180 MHz |
| 9 | | Fifo buffer for 134 M events |
| 10 | | USB 3.0 to host PC |
| 11 | | Reference clock output |

FIG 1. Block diagram of the instrument architecture.

A divider can be inserted in the synchronization channel if high sync frequencies are being used; otherwise it can be bypassed. The timing units are locked to the same crystal clock, so that their time measurements are always synchronized and equally calibrated. The event timings may therefore be regarded as precise (picosecond) wall clock readings, no matter which input they come from. Therefore, the subtraction or comparison of event times is valid within and across all channels. Because of the TDC based time measurement, the time difference between events (e.g., START and STOP) can be arbitrarily long while still being determined at the full resolution. Relative error is determined only by the crystal and clock distribution characteristics. Another feature resulting from choosing the independent channels design is the ability to introduce arbitrary offsets in each channel with picosecond resolution, thereby completely eliminating the need for cable delay adjustments as known from traditional systems using Time-to-Amplitude Converters (TAC).

**C. Data acquisition schemes**

A conventional histogramming mode supports classic TCSPC applications such as fluorescence lifetime measurements. This is implemented in FPGA hardware so that an average event rate of 78 Mcps can be processed in each timing channel. The front end FiFo buffers mentioned previously permit handling the full input rate of $1/T_d$ for bursts of up to 2048 events. If there is no space in the FiFo, events will be dropped. This is signaled in the form of a software flag available to the host computer.

While the start-stop time spans are in theory unlimited in length, the histogramming mode is limited to 65536 time bins due to practical limits of the histogram storage. The bin width can also be increased (repeated doubling of the 80 ps base resolution) in order to increase the overall time span. Since the dead time is acting only within each channel, the histogrammer can process multiple photons per excitation/emission cycle at slow excitation rates. Hence, it can collect data much more efficiently than a TAC based system, which always needs to wait for the next stop event.

Apart from the classic histogramming mode, the system provides two time-tagging data acquisition modes introduced previously.[28] The difference between the two time-tagging modes (T2 mode and T3 mode) lies primarily in the handling of sync events from, e.g., a pulsed laser. In T2 mode, all timing signal inputs are functionally identical. Usually all inputs are used to connect photon detectors. This may include the sync input and the sync divider is then bypassed. The events from all channels are recorded independently and treated equally. In each case an event record is generated that contains information about the channel it came from and the arrival time of the event with respect to the overall measurement start. The timing is recorded with 80 ps resolution. Each T2 mode event record consists of 32 bits carrying the channel number and a time-tag. If the time tag overflows, a special overflow marker record is inserted in the data stream so that, upon processing of the data stream in the host computer, a theoretically infinite time span can be recovered at full resolution. Autocorrelations within a channel can therefore be calculated at the full resolution, but only starting from lag times larger than the dead time. However, dead times, including those of the detectors, exist only within each channel but not across the channels. Therefore, cross correlations can be calculated down to zero lag time.

Exactly as in histogramming mode the event timing records from the TDCs are first queued in separate FiFo buffers per channel, each capable of holding up to 2048 event records. The FiFo inputs are fast enough to accept records at the full speed of the time digitizers (up to 1.53 GHz). This means that even during fairly intense bursts of photons, provided that there is space in the FiFo, no events will be dropped except those possibly lost due to the dead time. If there is no space in the FiFo, events will be dropped. This is signaled in the form of a software flag available to the host computer.

After temporal sorting, the T2 records from all channels are queued in another FiFo buffer capable of holding up to 134,217,728 event records. This FiFo is continuously read by the host PC, thereby making room for new incoming events. Even if the average read rate of the host PC is limited, bursts with much higher rate can be recorded for some time. A FiFo overrun can occur only if the average count rate over a longer time period exceeds the readout speed of the PC. Then the measurement must be aborted because data integrity cannot be maintained. However, on a reasonably modern PC (Intel Core i7 4770 3.6 GHz, Windows 10), sustained average count rates over 80 Mcps were obtained. This total transfer rate must be shared by all input channels. For most practically relevant applications, the effective rate per channel is expected to be sufficient.

In T3 mode, the sync channel is connected to a typically periodic synchronization signal, usually from an excitation source. As far as the experimental setup is concerned, this is the same as in TCSPC histogramming mode. The main objective of T3 mode is to allow using high sync rates from mode-locked lasers which would swamp the data stream if they were handled as in T2 mode. In order to enable sync rates > 78 MHz, the new TCSPC electronics provide a sync divider as introduced previously.[28] This permits sync rates as high as 1.2 GHz.

The event records in T3 mode are composed of two timing figures: 1) the start-stop timing difference between the photon event and the last sync event and 2) the arrival time of the event pair on the overall experiment time scale (the time tag). The time tag is obtained by counting sync pulses. From the T3 mode event records, it is therefore possible to precisely determine which sync period a photon event belongs to. Furthermore, since the sync period is also known precisely, the arrival time of the photon with respect to the overall experiment time can be reconstructed.

Each T3 mode event record consists of 32 bits carrying the channel number, the start-stop time, and the sync count. If the sync count overflows, a special overflow marker record is inserted in the data stream, so that upon processing of the data stream a theoretically infinite time span can be recovered. 15 bits of the record are used for the start-stop time difference, covering a time span of $32768 \times R$, where $R$ is the chosen resolution. At the best possible resolution of 80 ps this results in a time span of 2.62144 µs. If the time difference between a photon and the last sync event is larger, the photon event cannot be recorded. This is the same as in histogramming mode, where the number of bins is also finite. However, by choosing a suitable sync rate and a compatible resolution $R$, it is possible to reasonably accommodate all relevant experimental scenarios. $R$ can be chosen in doubling steps between 80 ps and 335.5 µs. The data transfer uses the same FiFo concept as in T2 mode. Again, sustained average count rates over 80 Mcps can be obtained. In T3 mode, the full transfer rate is available for the detector input channels since the sync events are handled implicitly.

The throughput limit above 80 Mcps in the time tagging modes are due to bandwidth limitations of the USB 3.0 interface. It corresponds to a throughput of 320 Mbytes/s which is close to exploiting the practical limits of USB 3.0 in the presence of overhead at the various levels of hardware and software.

## III. FUNDAMENTAL PERFORMANCE TESTS

### A. Timing precision

The current implementation of the described new TCSPC and time tagging system was used in various measurement scenarios to verify functionality and timing accuracy. Although the digital resolution of the TCSPC electronics is fixed at 80 ps by the TDC design, a timing uncertainty (jitter) due to noise is always present. This is true even for digital signals because of the finite slope of any signal transition. In order to test this quantity for the prototype design, the following test was performed. A test generator (CG635, Stanford Research Systems) provided pulses of 5 MHz repetition rate with transition times of 370 ps (10 to 90%). The steep transitions ensure that the time measurement results are only insignificantly influenced by noise, as shown by earlier measurements on TDCs with substantially higher resolution and precision.[39] This signal was fanned out through a reflection free splitter so that five identical signals were obtained. These were fed to the sync input and four input channels of the device under test using approximately identical cable lengths. The device was then operated in histogramming mode, where the time differences between sync and the respective input channel are recorded. The result is shown in figure 2. Each peak at 10, 20, 30, and 40 ns represents the histogram obtained for one channel. Using the software adjustable offset of each channel, the 10 ns spacings between the peaks were set arbitrarily for best view. Numerical analysis of the distributions shows that the r.m.s. timing jitter is typically 80 ps. It should be noted that this is the overall error including the sync channel and the respective detector channel. The single channel measurement jitter would correspond to 80 ps / $\sqrt{2}$ which gives about 56 ps.

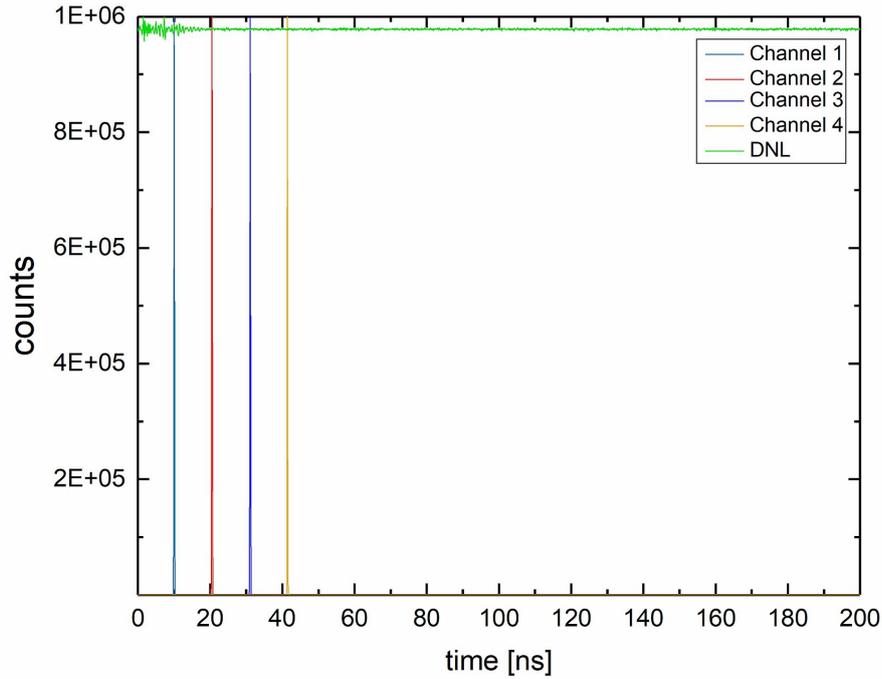

FIG 2. Constant delay timing histograms of channels 1..4 vs. sync input for assessing timing jitter (sharp peaks at about 10, 20, 30 and 40 ns). The x-axis offsets of the individual histograms were arbitrarily chosen for clarity of display only. The mostly flat line at the top is the result of a DNL measurement, i.e. the TCSPC histogram of uncorrelated signals.

In order to verify the precision of the White Rabbit remote synchronization for our own implementation, the following test was conducted: Two of the new time taggers were remotely clock-synchronized over a White Rabbit fiber link of 5 km on spool. The generator signal as described above was then split and fed to one input of either time tagger device. The r.m.s. timing uncertainty was then determined as for the local measuremet on a single device. This resulted in an r.m.s. jitter of 89 ps. This is slightly more than the timing jitter across the local channels but very good for a remote measurement over 5 km distance. Applying error propagation laws the r.m.s. jitter contribution of the White Rabbit link can be calculated, which results in 39 ps.

**B. Differential nonlinearity**

An intrinsic system property with high relevance, e.g., for fluorescence lifetime measurements is differential nonlinearity (DNL). It essentially describes the systematic error of the time digitizer's bin widths. In order to quantify this, another standard experiment was performed. A test generator (CG635, Stanford Research Systems) providing pulses at a rate of 5 MHz was used as the sync signal. A second generator (DS1090, Dallas Semiconductor) provided a 1 MHz pulse train with 8% dithering in frequency which was amplified and fed to one of the detector input channels. The use of independent generators, with one of them additionally dithering its period, ensured that from the perspective of the device under test, the signals were effectively uncorrelated. Therefore, the expected histogram of time differences should be evenly filled. Any deviation from a flat line would be due to either residual error of counting statistics or systematic error representing DNL. In order to minimize the residual error of counting statistics (within reasonable limits of measurement time), the experiment was allowed to run until the largest count had reached one million. The result is shown as the mostly flat line at about 1e6 in figure 2. Numerical analysis results in an r.m.s. deviation of 0.5% from the average. Inspecting the histogram in more detail, one notices that the part above 15 ns is very flat while the section up to 15 ns shows some ripple with a peak-to-peak value of about 5%. This is due to crosstalk among the input comparators leading to small shifts in the measured arrival times of the input channel when the sync input switches. Nevertheless, the effective DNL is by an order of magnitude better than that of many commercial TDC designs where DNL values as high as 50% or even 100% are not uncommon.

**C. Fluorescence lifetime measurements**

Apart from the fundamental tests with generator signals, fluorescence lifetime measurements were performed with dyes having known lifetimes. Figure 3 shows a fluorescence lifetime measurement of coumarin 6 in ethanol.

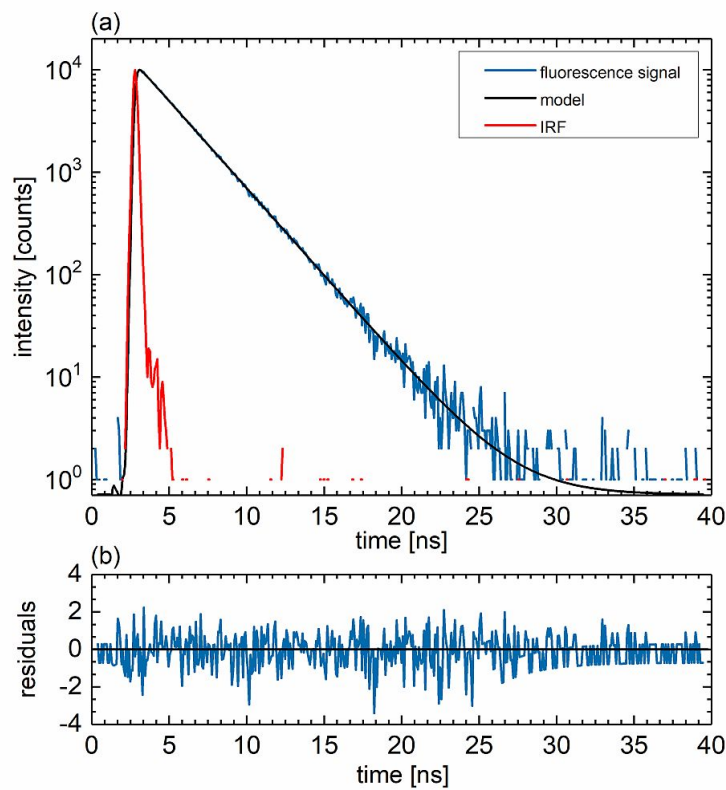

FIG 3. Fluorescence lifetime measurement of coumarin 6 in ethanol and fit results. Top panel: fluorescence decay (blue), model (black), and IRF (red). Lower panel: fit residuals.

Data were collected with a FluoTime 300 fluorescence lifetime spectrometer (PicoQuant, Berlin, Germany) using a PMA Hybrid 42 detector module (PicoQuant, Berlin, Germany) and the new timing electronics (MultiHarp 150, PicoQuant, Berlin, Germany). The sample was excited at 465 nm (LDH 470, PicoQuant, Berlin, Germany) and fluorescence detected at 510 nm. The Instrument Response Function (IRF) for iterative reconvolution was recorded using a Ludox dispersion in water. A fitted lifetime of 2.55±0.01 ns with an excellent $\chi^2_{red}$ of 1.0020 was obtained. The lower panel of the figure shows the fit residuals.

**D. Correlation measurements**

Another typical application of high resolution time tagging is the measurement of second-order correlations, g$^{(2)}(\tau)$, in quantum optics. Figure 4 shows the results of such a measurement.

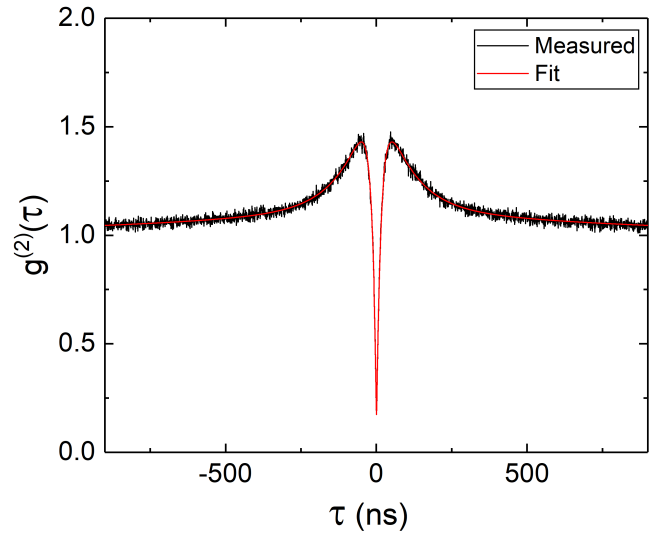

FIG 4. Second-order correlation g$^{(2)}(\tau)$ of the fluorescence signal from a single nitrogen-vacancy color center in diamond (black), fitted with a four-level model (red).

The raw data was recorded by detecting the fluorescence signal from a single nitrogen-vacancy color center in diamond using an home-built confocal setup[2]. The color center was excited with a continuous wave 532 nm laser, which was focused onto the sample using a NA = 1.35 objective lens. The collected light was spectrally filtered by a dichroic mirror and a 650 nm longpass filter and spatially discriminated by a pinhole before it is sent to two avalanche photodiodes in a Hanbury-Brown and Twiss configuration. The photon arrival times were recorded with the new timing electronics in T2 mode with a timing resolution of 80 ps. The correlation data can be fitted well by a four-level model[42] from which a value of g(2)(0) = 0.13 ± 0.01 < 0.5 could be obtained, proving the quantum emission from a single emitter.

---

[2] Courtesy of Florian Böhm, Tim Kroh and Oliver Benson, Nano Optics Group, Humboldt-Universität zu Berlin.

## IV. HIGH SPEED AND SPECTRALLY RESOLVED FLIM EXPERIMENTS

### A. Motivation

The purpose of this section is to present an exemplary, representative life science application using the new time tagging system, which overcomes several problems associated with multi-channel TCSPC systems and provides added benefit of faster data acquisition due to the instrument's capability to operate at high sustained data throughput independently in all channels. While this could be demonstrated in regular FLIM measurements, an even more interesting application scenario is spectrally resolved FLIM (sFLIM). It is well known that sFLIM is an excellent tool to quantitatively separate constituent fluorescence signals from different fluorescent species in a complex biological sample, as shown by Niehörster et al.[43] Such requirements exist in real medical research scenarios, e.g., the distinction of auto-fluorescent species in tissue sections.

### B. Experimental Setup

The confocal setup used for FLIM measurements in this study is shown in figure 5.

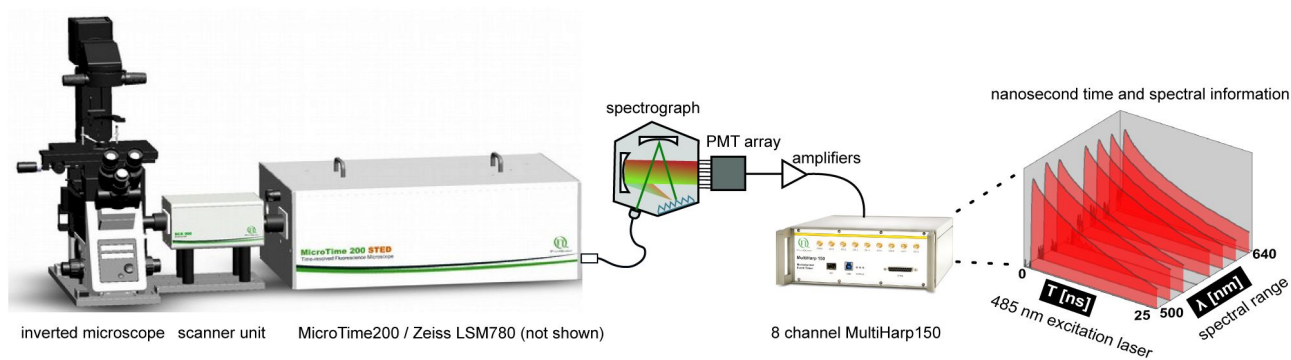

FIG 5. Schematic illustration of sFLIM setup used for experimental study.

Measurements with dye solutions were performed on a confocal time-resolved microscope (MicroTime 200, PicoQuant) equipped with a galvo scanner (FLIMbee, PicoQuant). Tissue sections were investigated using a Zeiss LSM 780 equipped with an sFLIM Upgrade Kit (PicoQuant). For efficient sample excitation, a pulsed laser with a wavelength of 485 nm (LDH-D-C-485, PicoQuant) was used and operated at a repetition rate of 20 MHz. To efficiently separate excitation from fluorescence signal, a major dichroic (zt488, AHF Analysentechnik AG, Germany) was placed in the reflected beam path after sample excitation. A long pass filter with cut-off at 500 nm (500/LP, AHF Analysentechnik AG, Germany) served to suppress scattered excitation laser light. For confocal imaging, the pinhole diameter was set to 100 μm. To spectrally separate the emitted fluorescence signal into its constituent wavelengths components, a detection system consisting of a spectrograph (Shamrock SR-163, Andor Oxford Instruments, UK) and a 16-channel PMT array with GaAsP (gallium arsenide phosphide) cathodes (H13123-40, Hamamatsu, Germany) was custom built by PicoQuant. Spectral splitting was achieved by using a grating with 600 l/mm and 500-nm blaze (SR1-GRT-0600-0500, Andor Oxford Instruments, UK). Corresponding to the eight TCSPC channels, the system was arranged to create eight spectral channels covering a range from 490 nm to 640 nm with spectral steps of 18.8 nm between successive channels. The signal from each detector was amplified by an individual amplifier module (PAM 102-P, PicoQuant) and connected to the corresponding timing input of the TCSPC electronics. The detector was operated at 1kV. For dye measurements in aqueous solution, the beam was focused into the sample through a 60x/1.2 NA water objective (UPlanSApo, Olympus USA). Human lung tissue samples fixed on cover slides were imaged using a plan-apochromat 63x/1.40 DIC M27 oil-immersion objective (Carl Zeiss, Germany). Regions of interests (ROI) having a size 80 μm x 80 μm (512 x 512 pixels) were imaged with a pixel dwell time of 1.5 μs. The whole setup was operated with the commercially available software package SymPhoTime 64 (PicoQuant), which was also used for data analysis.

## C. Performance of lifetime measurements at high count rates

A first experiment was performed to demonstrate the high photon throughput capability of the TCSPC electronics with simultaneous data collection in synchronized but independent and spectrally resolved detection channels. The measurement sample consisted of a diluted solution of Atto Rho6G dye (ATTO-TEC, Germany) in distilled water at a concentration of 100 µM. Measurements were performed at different laser excitation power to cover a detection throughput ranging from 1 Mcps to 114 Mcps (sum of all spectral channels). During these sequential measurements, the laser power before the objective was measured to be in the range of 0.1 µW to 15 µW. At very high count rates, data could only be acquired for limited time periods as the sustained data throughput limit of the acquisition software was reached. Emission spectra for each count rate were normalized with respect to count rate in the maximum emission channel (~565 nm). As can be seen in figure 6, obtained emission spectra are in qualitative agreement with data obtained from the literature, even at high count rates.[44]

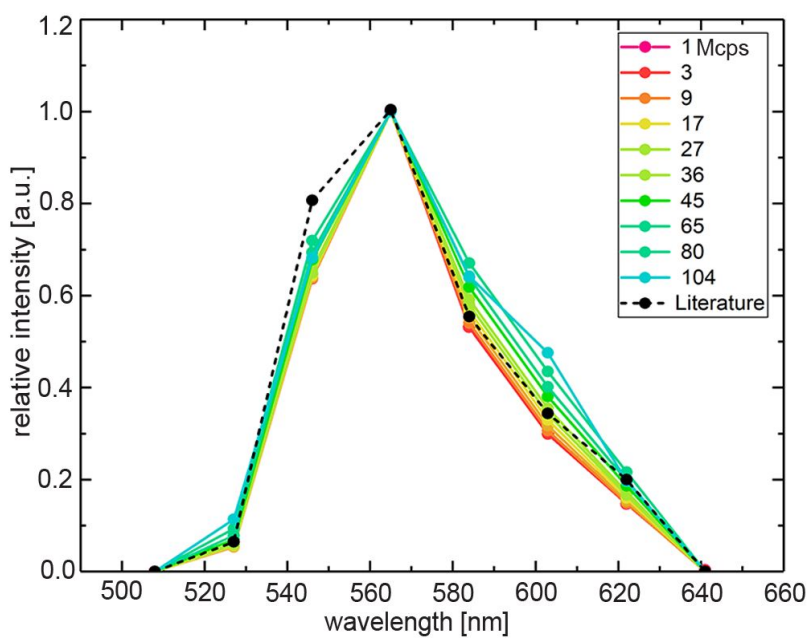

FIG 6. Normalized intensity vs. wavelength for measured emission spectra of Atto Rho6G dye solution at different detection photon count rates.

The deviations can be explained by the fact that the instrument's sensitivity was not calibrated between the 8 spectral detection channels. In some spectral channels, a small intensity deviation from low count rates towards high count rates was observed. This deviation is due to the detector pulse-pile-up effect, which leads to losses of detected photon counts at very high count rates. Fluorescence lifetime analysis was performed to verify whether high detection rates resulted in any significant changes in the lifetime properties of the freely diffusing dye. A bi-exponential model was fitted to the recorded fluorescence decays to obtain lifetimes for each independent channel. Pulse pile-up is a very well-known effect resulting from the detector physics which directly affects the photon count rates. The effect is caused primarily by the temporal width of the individual detector pulses (~1 ns). At high count rates, the photon statistics imply that the probability is quite high for photon emissions closer than the individual detector pulse width. This means that successive detector pulses may overlap and merge into one detection pulse. Since the TCSPC electronics cannot discriminate between successive original pulses, they will be registered as one photon event. This results in a special kind of histogram distortion where early photons are underrepresented. A correction scheme for the systematic errors resulting from this type of histogram distortion was already established in our previous work.[45] It physically models the photon losses due to detector pulse pile-up and incorporates the loss in the decay fit model employed to obtain fluorescence lifetimes and relative amplitudes of the decay components. A key parameter for the photon loss model is the closest pulse spacing in time that a particular combination of detector and TCSPC electronics can resolve. This parameter (called Delta-Pulse in SymPhoTime 64) was also included in the reconvolution fitting model used for analysis of the fluorescence decays. The parameter was set to 2.8 ns to yield optimal fit quality for the detector in use. Results from the bi-exponential fitting procedure are shown in figure 7. At an observed overall count rate below 50 Mcps, a lifetime of 3.96 ± 0.01 ns (mean ± SD of all measurements at all channels) was obtained, which is in good agreement with reported lifetime of Atto Rho6G.[46]

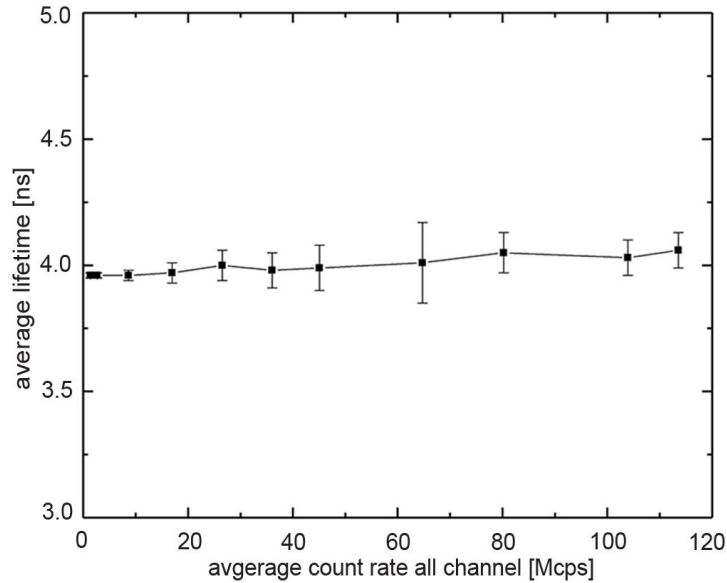

FIG 7. Fluorescence lifetime measurement of Atto Rho6G in water at different photon count rates summed across all detection channels. Obtained data was fitted taking into account an estimated instrument response function and detector pulse pile-up effects. Quantification of fluorescence lifetime is given as mean value for different count rates in all detection channels. The standard deviation of the mean values is shown by error bars.

At count rates as high as 114 Mcps, the fitting model still performed well, but a slight increase in lifetime (0.1 ns) could be observed.

Furthermore, we also demonstrate that it is possible to perform measurements and obtain reliable results beyond the classic pile-up limit. After summing the data across all spectral channels, it was possible to reach count rates up to 114 Mcps. At the chosen excitation rate of 20 MHz, more than 5 fluorescence photons were collected during one excitation cycle, still consistently performing fluorescence decay fitting including the pulse pile-up correction to obtain reliable lifetime values. Compared to the classic TCSPC pile-up limit (at maximum 10% of the excitation rate), measurements could now be performed 50 times faster. This high throughput detection capacity is made possible by the short dead-time of the new TCSPC electronics and the splitting of the fluorescence into several detection channels, as well as appropriately accounting for pulse-pile up and dead time effects in the fitting model during analysis.

**D. Applicability and benefit for life science research**

Here, we present a representative, exemplary life science application of the new time tagging system, which overcomes several problems associated with multi-channel TCSPC systems and provides the benefit of faster data acquisition speeds due to the instrument's capability to operate at high sustained data throughput independently in all channels. First, to demonstrate the advantage gained in terms of acquisition speeds, two separate measurements were performed using the experimental setup depicted in figure 5. The first set of measurements consisted of detecting count rates below 10% of the laser repetition rate of 20 MHz (i.e. 2 Mcps) in the spectral channel corresponding to emission maximum for excitation at 485 nm. This we term as classic sFLIM measurement. In the second set of measurement the laser power was increased to obtain significantly higher photon throughput, way above the classic TCSPC detection limit. This measurement mode shall be referred to as rapid sFLIM. A laser power of 2 µW and 15 µW was measured after the objective for classic and rapid sFLIM measurements, respectively. By setting the measurement stop condition to 1500 photons in the brightest pixel of the imaged Region Of Interest (ROI), the time taken to reach this stop condition for these two sFLIM acquisition modes were compared. It was observed that, at comparable levels of accuracy, the rapid sFLIM measurements were six times faster than the classic mode of acquisition. Furthermore, during rapid sFLIM acquisition, it was possible to reach count rates of 65 Mcps (summed over all channels) in the brightest pixel compared to 11 Mcps under the classic sFLIM operation regime. Next, the quality of lifetime fitting was evaluated by analyzing fluorescence decays obtained at different count rates for different species present in the sample. For this purpose, as shown in figure 8, ROIs from three distinct autofluorescent species, namely Red Blood Cells (RBCs), collagen and alveolar macrophages, present in human lung tissue samples were selected and a bi-exponential model was fitted to the recorded decays.

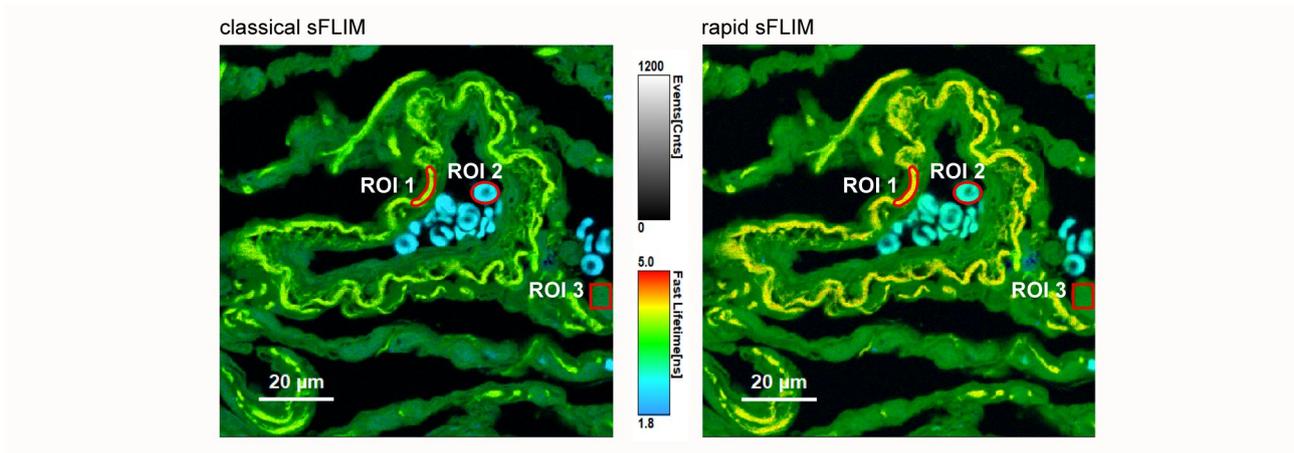

FIG 8. Representative mean photon arrival time confocal FLIM image from classic sFLIM (left) and rapid sFLIM (right) acquisition of highly autofluorescent human lung tissue sample. Fluorescence decays obtained from three distinct autofluorescence species were fitted to obtain average fluorescent lifetime values (see Table 1). ROI 1, 2, and 3 (marked red) correspond to collagen, Red Blood Cells (RBCs) and Alveolar Macrophages (AM).

The fitting model, as outlined above and published previously[45], also included the pulse pile-up correction to the fluorescence decays obtained from these regions using the two modes of FLIM acquisition, as explained above. Results from this lifetime analysis are shown in Table 1.

| ROI | Autofluorescent Specie | Lifetime ± std. dev. [ns] | |
|---|---|---|---|
| | | classic sFLIM | rapid sFLIM |
| 1 | Collagen | 3.6 ± 0.01 | 3.7 ± 0.01 |
| 2 | RBCs[a] | 2.4 ± 0.02 | 2.5 ± 0.01 |
| 3 | AM[b] | 2.7 ± 0.01 | 2.8 ± 0.02 |

**Table 1:** Quantification of lifetime for three distinct autofluorescent species in human lung tissue. (mean ± std. dev. from three independent measurements). [a]RBCs: red blood cells  [b]AM: alveolar macrophage

A small lifetime increase of 0.1 ns throughout all regions for the rapid sFLIM measurement was observed compared to classic FLIM. This increase in lifetime values corresponds to the increase in lifetimes obtained from the dye in solution measurements at very high count rates. In figure 8, we notice that some of the regions which appear green in classical sFLIM image show up as yellow in rapid sFLIM image i.e. lifetimes in rapid sFLIM appear longer than classical sFLIM image. To explain these differences, histograms of lifetime distributions for imaging data from classical and rapid sFLIM imaging (figure 8) are shown in figure 9, obtained using two different methods: (a) fast lifetime (i.e. calculated from the mean photon arrival time in a pixel; no correction applied) and (b) reconvolution fitting model applying a correction scheme for pulse pile-up effects observed for higher detection count rates (the rapid sFLIM correction). The differences in lifetime of the two FLIM images depicted by the color of autofluorescent species shows up as a shift in the two lifetime histograms obtained by mean photon arrival time calculation without any correction for the pulse pile-up effect. However, lifetime histograms obtained after using bi-exponential reconvolution fitting model by taking into account the rapid sFLIM correction displays a smaller shift between peaks corresponding to classical and rapid sFLIM measurements (figure 9b). Thus, the rapid sFLIM correction allows for a more correct lifetime calculation for FLIM measurements at very high count rates.

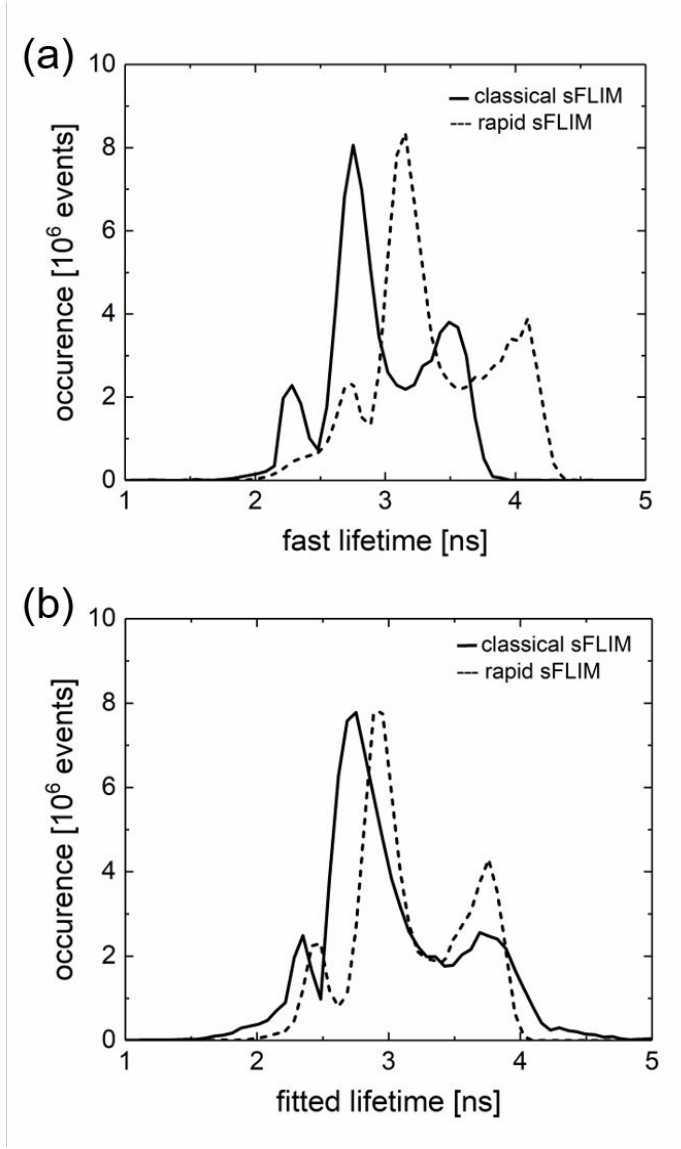

**FIG 9.** Representative histograms of lifetime distributions for imaging data obtained from classical sFLIM and rapid sFLIM measurements of human lung tissues sample (see figure 7). Lifetime values were obtained using two different methods: (a) fast lifetime (i.e. calculated from the mean photon arrival time in a pixel) and (b) reconvolution fitting model applying a correction scheme for pulse pile-up effects observed for higher detection count rates.

## V. CONCLUSION AND OUTLOOK

We showed that the speed limitations of TCSPC can be lifted by minimizing the dead time of the timing electronics while maintaining high temporal resolution. Similarly, the overall throughput can be increased by means of simultaneous readout of many detector channels through independent

TCSPC channels. We described a new time tagging system design, providing up to 16 input channels, a dead time of only 650 ps, very high time tagging throughput, and a timing resolution of 80 ps. In order to facilitate remote synchronization of multiple such instruments at the same high timing precision over kilometer distances, the TCSPC electronics provide an interface for White Rabbit fiber optic networks.

Experimental benchmark results such as timing precision and DNL of the instrument were shown. In an applied scenario, we demonstrated how the new time tagging system design enables high speed fluorescence lifetime measurements, overcoming the classic pile-up limit and permitting a 6-fold increase in fluorescence photon throughput. Employing an advanced decay fit model incorporating corrections for detector pulse-pile-up, it was possible to obtain accurate lifetimes even with count rates as high as 65 Mcps, apart from a small systematic error. Likely origins of the discrepancy may lie in the detector itself as well as in the way closely spaced detector pulses shift the baseline level in the electrical signals. Further investigations in this matter are under way.

In addition to the basic characterizations, we showed application results from spectrally resolved and high speed fluorescence lifetime imaging in biomedical research. It is well known that sFLIM is an excellent tool to quantitatively separate constituent fluorescence signals from different species in a complex biological sample. However, it requires high speed data acquisition. A historical obstacle of classic FLIM measurements in general, and sFLIM measurements in particular, is that it was relatively slow compared to intensity-based measurements. This was primarily due to the dead time constraints of the TCSPC hardware. We overcome this limitation by combining low dead time TCSPC electronics, operating independently and in parallel for each spectral channel, a highly sensitive PMT detector array with GaAsP cathodes and a new algorithm for correcting the pulse pile up of the detector. This broadens the applicability of rapid sFLIM measurements, notably in the fields of material as well as life sciences. In the near future we expect to improve the timing resolution of the system without compromising the short dead time.

**ACKNOWLEDGEMENTS**

The authors thank Paja Reisch, Max Tillmann, Matthias Patting and Marcus Sackrow for help with characterization measurements and debugging as well as André Devaux for proof-reading and corrections. Our gratitude for the second-order correlation measurement goes to Florian Böhm, Tim Kroh and Oliver Benson of the Nano Optics Group at Humboldt-University Berlin. We are also grateful to Katharina Hellwig for excellent assistance with lung tissue sample preparation.

We acknowledge financial support by the European Fund for Regional Development of the European Union in the framework of project iMiLQ, administrated by Investitionsbank Berlin as part of the Program to Promote Research, Innovation, and Technologies (ProFIT) under grant number 1015946. The application work was supported by funding from the European Union's Framework Program for Research and Innovation Horizon 2020 (2014-2020) under the Marie Skłodowska Curie Grant Agreement No. [675332]. Additionally, the work was supported by the German Research Foundation (DFG SFB-TR84) to A.C.H (Z1a).